\documentclass{elsart4-1}

\usepackage{epsfig}

\usepackage{amssymb,amsmath}
\usepackage{cite}

\usepackage[english,francais]{babel}

\newtheorem{e-proposition}[theorem]{Proposition}

\newtheorem{e-definition}[theorem]{Definition\rm}

\renewcommand{\theequation}{\arabic{equation}}
\def\og{\leavevmode\raise.3ex\hbox{$\scriptscriptstyle\langle\!\langle$~}}
\def\fg{\leavevmode\raise.3ex\hbox{~$\!\scriptscriptstyle\,\rangle\!\rangle$}}

\begin{document}

\def\ra{\rangle}
\def\la{\langle}
\def\be{\begin{equation}}
\def\ee{\end{equation}}
\def\CO{\mathcal{O}}
\def\CM{\mathcal{M}}
\def\CN{\mathcal{N}}
\def\CV{\mathcal{V}}
\def\CT{\mathcal{T}}
\def\CL{\mathcal{L}}
\def\x{\xi}
\def\Tr{\rm Tr}
\def\s{{\sigma}}
\def\rh{{\rho}}
\def\t{{\tau}}
\def\tt{{\tilde{\tau}}}
\def\l{{\lambda}}
\def\D{\Delta}
\def\m{\mu}

\begin{frontmatter}


\selectlanguage{english}
\title{Minimal String Theory\thanksref{talk}}
\thanks[talk]{Talk presented by N.S. at Strings '04, June 28-July 2, Paris}




\selectlanguage{english}
\author[seiberg]{Nathan Seiberg}
\ead{seiberg@ias.edu}
\author[shih]{David Shih}
\ead{dshih@princeton.edu}

\address[seiberg]{School of Natural Sciences, Institute for Advanced Study,
Princeton, NJ 08540 USA}
\address[shih]{Department of Physics, Princeton University, Princeton, NJ
08544 USA}

\begin{abstract}
We summarize recent progress in the understanding of minimal
string theory, focusing on the worldsheet description of physical
operators and D-branes. We review how a geometric interpretation
of minimal string theory emerges naturally from the study of the
D-branes. This simple geometric picture ties together many
otherwise unrelated features of minimal string theory, and it
leads directly to a worldsheet derivation of the dual matrix
model.

\vskip 0.5\baselineskip


\keyword{D-branes; 2D Gravity; Matrix Models}
\endkeyword
\end{abstract}
\end{frontmatter}


\selectlanguage{english}

\section{Introduction}

Minimal string theories are an important class of tractable,
exactly solvable toy models. Despite their simplicity, they are
interesting laboratories for the study of string theory, because
they contain many of the desirable features of critical string
theory, including D-branes, holography and open/closed duality.
These general phenomena are realized concretely in minimal string
theory through the well-known duality with large $N$ random matrix
models. (For reviews and older references, see e.g.\
\cite{GinspargIS,DiFrancescoNW}.) This duality is special and
particularly interesting because both dual theories -- the
pertubative minimal string and the matrix model -- are solvable.

Recent progress in the study of Liouville theory
\cite{DornSV,TeschnerYF,ZamolodchikovAA,FateevIK,TeschnerMD,ZamolodchikovAH,PonsotNG}
has spurred a renewed interest in the minimal string (see e.g.\
\cite{McGreevyKB,MartinecKA,KlebanovKM,McGreevyEP,KlebanovWG,SeibergNM,GaiottoYB,HanadaIM,KutasovFG,AmbjornMY}),
leading to new insights, some of which we will review here. We
will summarize and highlight some new results which are described
in much more detail in \cite{SeibergNM,KutasovFG}. We will use
worldsheet techniques to derive the dual matrix model. Along the
way, a simple geometrical picture will emerge, which will serve to
unify many features of the minimal string.

For simplicity, we will focus on bosonic minimal string theories,
which are labelled by two relatively prime integers $p<q$. The
worldsheet sigma model consists of two parts: $(p,q)$ minimal CFT
and Liouville theory.\footnote{One can also consider type 0
minimal string theory, whose worldsheet description consists of
superminimal CFT coupled to $\CN=1$ super-Liouville theory.
Although much more complicated on the worldsheet, these models
turn out to be surprisingly similar to their bosonic cousins. In
particular, they have an analogous geometrical interpretation that
leads to a derivation of the matrix model. A detailed analysis of
these models is given in \cite{SeibergNM}.} In the next two
sections, we will describe these two worldsheet CFTs in detail,
and we will show how they are combined to form minimal string
theory. Section 2 focuses on the closed minimal string and
describes the spectrum of physical operators, while section 3
discusses the D-branes of minimal string theory. The main purpose
of these two sections is to collect a diverse list of facts about
minimal string theory. These facts are then tied together in
section 4, using an auxiliary Riemann surface $\CM_{p,q}$ that
emerges from the D-branes. In section 5 we show how the same
Riemann surface leads to a worldsheet derivation of the matrix
model. Finally, various conclusions are collected in section 6.

\section{Minimal String Theory on the Worldsheet}

\subsection{$(p,q)$ minimal CFT}

The first ingredient in the worldsheet recipe for minimal string
theory is $(p,q)$ minimal CFT
(for a review, see e.g.\ \cite{DiFrancescoNK}). The minimal models
are labelled by their central charge
\begin{equation}
\label{minc}
c = 1-{6(p-q)^2\over pq} < 1
\end{equation}
Unlike most CFTs, the minimal models have only a finite number of
primary operators. We will denote these operators by $\CO_{r,s}$
with $r=1,\dots,p-1$, $s=1,\dots,q-1$ and $
\CO_{p-r,q-s}=\CO_{r,s}$. Their conformal dimensions are given by
\begin{equation}
\label{primaries}
\Delta(\CO_{r,s}) = {(r q-s p)^2-(p-q)^2 \over 4p\,q}
\end{equation}
The formula for the dimensions implies that every primary
corresponds to a degenerate representation of the Virasoro
algebra. That is, they all have null states among their Virasoro
descendants. One can systematically exploit this fact to
completely constrain the multiplication table of these operators.
This results in a set of {\it fusion rules}, which take the form
\begin{equation}
\label{fusion}
 [\CO_{r_1,s_1}]  \times [\CO_{r_2,s_2}] =  \sum_{r_3=1}^{p}
 \sum_{s_3=1}^{q} N_{(r_1,s_1)(r_2,s_2)(r_3,s_3)} [\CO_{r_3,s_3}]
\end{equation}
where the fusion numbers $N_{(r_1,s_1)(r_2,s_2)(r_3,s_3)}$ are
either zero or one.
The notation $[\CO]$ in (\ref{fusion}) is meant to indicate the
primary operator $\CO$ and all its Virasoro descendants. In other
words, the fusion rules tell us how to multiply two Virasoro
representations, but they do not tell us the details of the actual
operator product coefficients.

\subsection{Liouville theory}

The other ingredient in the worldsheet construction of minimal
string theory is Liouville theory. We can think of this as a
theory of a scalar field in two dimensions with action
\begin{equation}
\label{LiouvilleS} S = {1\over4\pi}\int d^2z\,
\bigg((\partial\phi)^2-4\pi \mu e^{2 b \phi} \bigg)
\end{equation}
The parameter $b>0$ is called the Liouville coupling constant,
while $\mu$ is called the cosmological constant. We will assume
$\mu\ne 0$ throughout, and we will work in units where $\mu=1$ for
simplicity.

For conformal invariance, one also needs to include a ``background
charge"
\begin{equation}
\label{Qdef}
Q = b+{1\over b}
\end{equation}
which results in an asymptotically linear dilaton background at
$\phi\to -\infty$. The presence of nonzero $Q$ has several
effects. First, it shifts the central charge from the free field
value $c=1$ to
\begin{equation}
\label{Liouvillec}
 c=1+6Q^2 \ge 25
\end{equation}
Second, it changes the dimensions of primary operators $V_\alpha =
e^{2\alpha\phi}$ from $\Delta = -\alpha^2$ to
\begin{equation}
\label{Lprimaries}
\Delta(V_\alpha) = -\bigg({Q\over 2}-\alpha\bigg)^2 +{Q^2 \over 4}
\end{equation}
Representation theory of the Virasoro algebra with central charge
(\ref{Liouvillec}) tells us that the primary operators with
\begin{equation}
\label{Ldegenerate} 2\alpha_{r,s}={1\over b} (1-r) + b (1-s),
\qquad r,s\in \mathbb Z^+
\end{equation}
correspond to degenerate representations of Virasoro. As in the
minimal models, these operators have special fusion rules which
lead to a complete solution of Liouville theory
\cite{DornSV,TeschnerYF,ZamolodchikovAA}.

\subsection{Minimal string theory}

Now we can combine the two ingredients, together with the standard
ghosts, to form minimal string theory. We will refer to the
minimal CFT as the ``matter sector." Requiring the total central
charge of matter plus Liouville to be $c=26$ sets
\begin{equation}
\label{bpq}
b^2={p \over q}
\end{equation}
As we will see below, the fact that $b^2$ is rational leads to
many simplifications in minimal string theory.

Physical operators in minimal string theory are built out of the
operators of the Liouville, matter, and ghost sectors. As usual,
BRST invariance requires physical operators to have dimension
zero. However, in contrast with critical string theory, in minimal
string theory there are BRST invariant physical operators at all
ghost numbers \cite{LianGK}. This is a direct consequence of the
existence of degenerate operators in Liouville theory and the
matter sector. The most important operators for our purposes are
those at ghost number zero and one. Let us now describe them in
detail, starting with the operators at ghost number zero.

Clearly, by ghost number conservation, the ghost number zero
operators form a ring under multiplication by the OPE
\cite{Witten:1991zd}.  This ring is called the {\it ground ring},
and its elements take the form
\begin{equation}
\label{groundring}\begin{split}
 &\hat{\CO}_{r,s}=\CL_{r,s}\cdot \CO_{r,s}e^{2\,
 \alpha_{r,s}\, \phi },\qquad  1\le r\le p-1 \ , \quad 1\le s\le q-1
\end{split}\end{equation}
with $\alpha_{r,s}$ given by (\ref{Ldegenerate}) and $\CL_{r,s}$ a
certain polynomial in ghosts and Virasoro generators. Note that
the range of $r$ and $s$ means that there are exactly {\it twice}
as many ground ring elements as primaries in the minimal model.

The multiplication of the ring elements is constrained by the
fusion rules in the matter and Liouville CFTs. This allows us to
completely determine the ring relations, up to a few coefficients
which are justified later. In terms of the ring generators
\begin{equation}
\label{grfirst}\begin{split}
 &X \equiv {1\over2}\hat\CO_{2,1},\qquad Y\equiv {1\over2}\hat \CO_{1,2}
\end{split}\end{equation}
(note that $\hat\CO_{1,1}=1$) we find that
\begin{equation}
\label{grmult}
\hat \CO_{r,s}=U_{s-1}(X) U_{r-1}(Y)
\end{equation}
where the $U_{s-1}(X)$ are Chebyshev polynomials of the second
kind,
\begin{equation}
\label{chebyshevU}
U_{s-1}(X=\cos\theta) ={\sin s \, \theta \over \sin \theta}
\end{equation}
Since these polynomials are also the $SU(2)$ characters, their
products are the $SU(2)$ fusion rules. In particular, the
coefficients in this multiplication table are either zero or one.

The formula (\ref{grmult}) for the ring elements would obviously
be incorrect if it were not supplemented by additional relations
in the ring, since otherwise we would find infinitely many ring
elements. Comparing with the range of $r$ and $s$ in
(\ref{groundring}), we see that the correct ring relations to
impose are
\begin{equation}
\label{grrel}
 U_{q-1}(X)=U_{p-1}(Y) =0
\end{equation}
(With only $X$ present, this is familiar from the representation
ring of $\widehat{SU(2)}$.) The ring relations (\ref{grrel})
preserve the simplicity of the multiplication table, i.e.\ {\it
all the coefficients are still just zero or one}! In the
traditional worldsheet analysis of the minimal string, this simple
answer would arise as a surprising cancellation between
complicated OPEs of operators in the minimal CFT and Liouville
theory.

Having understood the ground ring and its structure in some
detail, we can now apply our knowledge to the study of the
operators at other ghost numbers. Ghost number conservation
implies that the set of all operators at a given ghost number form
a module under the action of the ground ring. The simplest such
module consists of the ghost number one ``tachyon" operators:
\begin{equation}
\label{tach}
 \CT_{r,s}= c\, \bar c\, \CO_{r,s} e^{2\beta_{r,s}\phi}, \quad
 1\le r\le p-1 \ , \quad 1\le s \le q-1
\end{equation}
BRST invariance requires $\Delta(\CT_{r,s})=0$, which according to
(\ref{primaries}) and (\ref{Lprimaries}) is satisfied
when\footnote{Note that $\beta_{r,s}$ is determined by a quadratic
equation, and we have chosen the root with $\beta_{r,s}<Q/2$. This
bound can be understood in the semiclassical approximation as a
requirement on the locality of the vertex operator
\cite{SeibergEB}.}
\begin{equation}
\label{betars}\begin{split}
 &2\beta_{r,s}= {p+q-|rq-s p|\over \sqrt{p\, q}}\cr
\end{split}\end{equation}
Unlike the ground ring elements (but like the minimal model
primaries), the tachyons satisfy a reflection relation:
\begin{equation}
\label{reflrel}
\CT_{p-r,q-s} = \CT_{r,s}
\end{equation}
Thus there are exactly as many independent tachyons as primaries
in the minimal model.

Since the tachyons form a module under the action of the ground
ring, a trivial application of the fusion rules in the matter CFT
leads to the following simple formula for the tachyons in terms of
the ring elements:
\begin{equation}
\label{tachmodii}
\CT_{r,s} =\hat \CO_{r,s}  \CT_{1,1} = U_{s-1}(X)
U_{r-1}(Y)\CT_{1,1}
\end{equation}
This formula is actually quite useful. We mention two
applications:

\begin{itemize}

\item The tachyons obviously do not form a faithful representation
of the ring, since there are half as many tachyons as ground ring
elements. Indeed, combining the reflection relation
(\ref{reflrel}) with (\ref{tachmodii}), we obtain a new relation
in the module:
\begin{equation}
\label{tachmodrel}
T_p(Y)-T_q(X)
= 0
\end{equation}
with $T_p(Y)$ the Chebyshev polynomials of the first kind,
\begin{equation}
\label{chebyshevT}
 T_p(Y=\cos \theta) = \cos p\, \theta
\end{equation}
The relation (\ref{tachmodrel}) can also be written as
$U_{p-2}(Y)-U_{q-2}(X)=0$.  This relation as well as (\ref{grrel})
were first discovered as relations in the fusion ring in
\cite{DiFrancescoZK}.

\item Using the ring and its module, we can easily derive some
correlation functions. For instance,
\begin{equation}
\label{tachcorr}\begin{split}
 \langle \CT_{r_1,s_1} \CT_{r_2,s_2} \CT_{r_3,s_3} \rangle &=   \langle   \hat{\CO}_{r_1,s_1}
    \hat{\CO}_{r_2,s_2}\hat{\CO}_{r_3,s_3}
     \CT_{1,1} \CT_{1,1} \CT_{1,1}\rangle \cr
    &=N_{(r_1,s_1)(r_2,s_2)(r_3,s_3)}
\end{split}\end{equation}
In other words, the three-point functions of tachyon operators in
$(p,q)$ minimal string theory are precisely the fusion rules of
the associated $(p,q)$ minimal CFT. This surprisingly simple
result had been derived previously using much more complicated
methods in \cite{DiFrancescoJD,GoulianQR,DiFrancescoUD}. Our new
derivation shows why the correlators are so simple: it is due to
the underlying simplicity of the ground ring.

\end{itemize}

\section{D-branes in Minimal String Theory}

In the previous section, we described how minimal string theories
are put together at the worldsheet level, focusing on the closed
string sector. Now let us turn to the open strings and describe
how to construct the D-branes of minimal string theory. These are
built out of the D-branes of Liouville theory and minimal CFT.
They are conveniently described using the boundary state
formalism, which associates to every D-brane a ``Cardy state"
labelled by a highest-weight Virasoro representation in the open
string channel.

In the minimal models, the Cardy states are in one-to-one
correspondence with the Virasoro representations, and therefore a
given minimal model has a finite number of branes given by the
number of primary operators. The branes are usually denoted by
$|k,l\rangle$, with $k$ and $l$ taking the same integer values as
for the primaries $\CO_{k,l}$ of the minimal model.

The D-branes of Liouville theory were discovered more recently
through the work of \cite{FateevIK,TeschnerMD,ZamolodchikovAH}.
There are two kinds of D-branes in Liouville theory. The first
kind, called FZZT branes \cite{FateevIK,TeschnerMD}, fall into a
continuous family parametrized by the ``boundary cosmological
constant'' $\mu_B$ which multiplies the boundary interaction
\begin{equation}
\label{deltaS}
 \delta \, S =\mu_B \oint e^{b\, \phi}
\end{equation}
Solving the equation of motion for $\phi$ on a worldsheet with
bulk action (\ref{LiouvilleS}) and boundary interaction
(\ref{deltaS}) gives rise to Neumann-like boundary conditions for
the Liouville field:
\begin{equation}
\label{fzztbc}
 \partial_\sigma\phi = -2\pi b\mu_B e^{b\phi}
 \end{equation}
The FZZT branes are extended semi-infinitely in $\phi$ space. One
way to see this is using the minisuperspace wavefunction of the
FZZT brane. In the semiclassical $b\to 0$ limit, this is given by
\begin{equation}
\label{minispace}
 \psi(\phi)= \langle \phi| \mu_B\rangle = e^{-\mu_B\,
 e^{b\, \phi}}
\end{equation}
According to (\ref{minispace}), the FZZT brane comes in from
$\phi=-\infty$ and dissolves at $\phi \approx - {1 \over b} \log
\mu_B$ (see figure 1).

In the boundary state formalism, the FZZT brane labelled by
$\mu_B$ corresponds to a Cardy state $|\,\sigma\rangle$ labelled
by the nondegenerate Virasoro representation with dimension
\begin{equation}
\label{Deltasigma}
\Delta={1\over 4} \sigma^2 +{Q^2 \over 4}
\end{equation}
(in the notation of (\ref{Lprimaries}), this corresponds to
$\alpha = {Q\over2}+{i\sigma\over2}$) where $\sigma$ and $\mu_B$
are related by\footnote{This relation can be motivated in
Liouville theory as the boundary analogue of the B\"acklund
transformation. In the bulk, the B\"acklund transformation maps
the Liouville field $\phi$ to a free field $\tilde\phi$. The FZZT
brane satisfies Neumann-like boundary conditions in $\phi$ space
(\ref{fzztbc}), which are mapped to purely {\it Dirichlet}
boundary conditions in $\tilde\phi$ space with
$\tilde\phi=\sigma$.}
\begin{equation}
\label{muBsigma}
 \mu_B= \cosh \pi \, b \,
\sigma
\end{equation}
This expression allows us to analytically continue $\sigma \in
\mathbb C$, which is a infinite multiple cover of $\mu_B\in
\mathbb C$.

The second kind of Liouville branes are the ZZ branes
\cite{ZamolodchikovAH}. These fall into a discrete, two-parameter
family parametrized by integers $m,n\ge 1$, and they are all
localized in the strong coupling region $\phi\to +\infty$. In the
boundary state formalism, the ZZ branes correspond to the
degenerate representations of Liouville theory, which according to
(\ref{Ldegenerate}) and the map $\alpha =
{Q\over2}+{i\sigma\over2}$ are given by
\begin{equation}
\label{sigmadegen}
 \sigma = \sigma(m,n) = i\bigg({m \over b} + nb \bigg),\qquad
 m,n\in \mathbb Z^+
\end{equation}
Subtracting the null vectors in the degenerate representation
leads to a formula for the boundary state of the ZZ branes in
terms of the FZZT branes \cite{ZamolodchikovAH}
\begin{equation}
\label{ZZbs}
|m,n\rangle = |\,\sigma(m,n) \rangle - |\, \sigma(m,-n)\rangle
\end{equation}
Notice that the two FZZT branes in (\ref{ZZbs}) have the same
value of the boundary cosmological constant \cite{MartinecKA}:
\begin{equation}
\label{samemuB}
 \mu_B = (-1)^m \cos \pi\,  n \, b^2
\end{equation}
We will give a geometric interpretation to this fact below.

\begin{figure}
\begin{center}
\epsfig{file=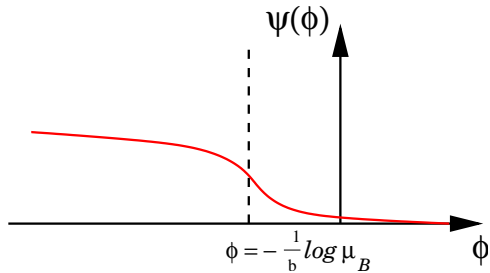, width=0.4\textwidth}
\end{center}
\caption{A plot of the minisuperspace wavefunction of the FZZT
brane. The dotted line denotes the location of the ``tip" of the
FZZT brane in $\phi$ space.} \label{fig:psi}
\end{figure}

Having described the branes of Liouville theory and minimal CFT,
we are now ready to form the D-branes of minimal string theory. We
simply tensor together a boundary state from Liouville theory
(either FZZT or ZZ) together with a boundary state $|k,l\rangle$
from the minimal model. (Note that we also have to set
$b=\sqrt{p\over q}$ in the formulas above.) However, not all of
these tensored boundary states are linearly independent. Before we
can write down the actual list of D-branes in minimal string
theory, there are three subtleties we have to take into account.

\begin{itemize}

\item Naively, it would seem that there are a number of different
FZZT branes and ZZ branes, distinguished by the choice of matter
state. However, it turns out that only the branes with matter
state $|1,1\rangle$ are independent.\footnote{This can be seen
from an analysis of the one-point functions. See \cite{SeibergNM}
for the details.} The FZZT branes with matter state $|k,l\rangle$
are related to those with matter state $|1,1\rangle$ by the
identifications
\begin{equation}
\label{fzztopfrelate}
\begin{split}
 | \,\sigma&; k,l\rangle =
\sum_{m'=-(k-1),2}^{k-1}\ \sum_{n'=-(l-1),2}^{l-1} |
\,\sigma+{i(m'q+n'p)\over \sqrt{pq}};1,1\rangle\cr
\end{split}
\end{equation}
The ZZ branes satisfy the same relations, thanks to (\ref{ZZbs}).
Note that these identifications are meant to be statements about
the BRST cohomology, i.e.\ the branes on the two sides of
(\ref{fzztopfrelate}) cannot be distinguished by any physical
observables.

\item Another consequence of the BRST cohomology is that not all
of the FZZT branes labelled by $\sigma$ are distinct. It turns out
that the independent FZZT branes are reduced to
$|\,\sigma\rangle \to |\,z\rangle$
with the parameter $z$ defined to be
\begin{equation}
\label{zdef} z = \cosh {\pi\sigma\over\sqrt{pq}}
\end{equation}
In the next section we will see that $z$ plays a central role in
the geometrical interpretation of minimal string theory.

\item Finally, there are subtleties having to do with Virasoro
representation theory for rational $b^2$. Without getting into the
details, let us just say that taking them into account (and
imposing the BRST cohomology) reduces the infinite family of
Liouville ZZ branes to a finite number $|m,n\rangle$ with $1\le
m\le p-1$, $1\le n\le q-1$, and $qm-pn>0$ \cite{SeibergNM}.

\end{itemize}

\medskip

\noindent Combining these three facts, we arrive at the final list
of independent D-branes in minimal string theory:
\begin{equation}
\label{Dbranelist}\begin{split}
 {\rm FZZT}:\quad &|\,z\rangle\otimes|1,1\rangle,\qquad\quad\,\,\, z\in \mathbb C\cr
 {\rm ZZ}:\quad &|\,m,n\rangle\otimes|1,1\rangle,\qquad 1\le m\le p-1, \quad  1\le
 n\le  q-1, \quad qm-pn>0
\end{split}\end{equation}

\section{Geometric Interpretation}

In the previous two sections, we collected many different facts
about open and closed minimal string theory. Now let us show how
these seemingly unrelated facts combine to form a simple geometric
picture of minimal string theory. The starting point is the
observation that the disk amplitude $Z(\mu_B)$ of the FZZT brane
is not a single valued function of $\mu_B$. Instead, if we define
\begin{equation}
\label{xydef}
x \equiv \mu_B = \cosh {\pi b\sigma} ,\qquad y \equiv
\partial_{\mu_B} Z(\mu_B) = \cosh {\pi\sigma \over b}
\end{equation}
then $x$ and $y$ satisfy the algebraic equation
\begin{equation}
\label{Fpq}
F_{p,q}(x,y) = T_p(y) - T_q(x) = 0
\end{equation}
This describes a genus ${(p-1)(q-1)\over 2}$ Riemann surface
$\CM_{p,q}$ with ${(p-1)(q-1)\over 2}$ pinched $A$-cycles
(singularities). An example of such a surface is shown in figure
2. The singularities occur at the simultaneous solution of
(\ref{Fpq}) and
\begin{equation}
\label{sings}\begin{split}
 &T_p'(y)=p\, U_{p-1}(y)=0\cr
 &T_q'(x)=q\, U_{q-1}(x)=0
\end{split}\end{equation}
We recognize (\ref{Fpq}) and (\ref{sings}) to be precisely the
ground ring relations and the relation in the tachyon module! Thus
the {\it algebraic} structure of the ground ring is directly
related to the {\it geometric} structure of the FZZT brane.

\begin{figure}[b]
\begin{center}
\epsfig{file=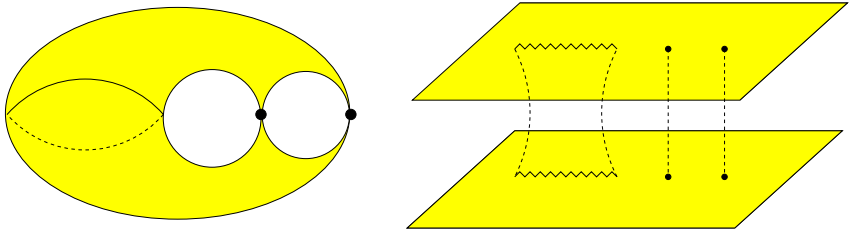, width=0.7\textwidth}
\end{center}
\caption{An example of a Riemann surface with pinched cycles,
viewed also as a double-cover of the complex plane.}
\label{fig:mpq}
\end{figure}

As promised, the parameter $z$ defined in (\ref{zdef}) plays an
important role in the geometrical description: {it is the \it
uniformizing parameter of $\CM_{p,q}$}. By this we mean the
following. It is trivial to see that the equation (\ref{Fpq}) for
the surface is solved by
\begin{equation}
\label{Fpqsolz} x=T_p(z),\qquad y=T_q(z)
\end{equation}
This means that, apart from the singularities, every point $(x,y)$
on the surface is in one-to-one correspondence with a point $z$ in
the complex plane. Thus, the complicated structure of $\CM_{p,q}$
is mapped to the complex $z$ plane by (\ref{Fpqsolz}), i.e.\ the
surface is {\it uniformized} by (\ref{Fpqsolz}). The singularities
are then points where this one-to-one correspondence breaks down.
For the surfaces described by (\ref{Fpq}), there are exactly two
values of $z$ corresponding to a given singularity.

Since the surface $\CM_{p,q}$ arose from the FZZT disk amplitude,
it is not surprising that it encodes the FZZT branes in a natural
way. What is surprising, however, is that it also knows about the
ZZ branes. Let us see how this comes about. Consider the following
one form on $\CM_{p,q}$:
\begin{equation}
\label{omegadef}
\omega\equiv y\, dx
\end{equation}
Then the D-branes correspond to line integrals of $\omega$. The
FZZT brane is obviously an integral of $\omega$ along an open
contour:
\begin{equation}
\label{fzztint}
Z(x) = \int_P^{x} \omega
\end{equation}
On the other hand, the $(m,n)$ ZZ brane is a difference between
two FZZT branes with the same value of $x=x_{m,n} = (-1)^m\cos{\pi
p\,n\over q}$, so it corresponds to an integral of $\omega$ along
a {\it closed} contour:
\begin{equation}
\label{zzint}
Z(m,n) = \oint_{B_{m,n}}  \omega
\end{equation}
This gives a geometric interpretation to the relation (\ref{ZZbs})
between the ZZ and FZZT boundary states.\footnote{The formula for
the ZZ brane as a closed contour integral was first derived for a
special case in \cite{KlebanovWG}.}

Note that since the disk amplitude $Z(m,n)$ is nonzero, the
contour of the $(m,n)$ brane must be a nontrivial cycle of the
surface. We can confirm this geometrically by noticing that
$B_{m,n}$ passes through $(x_{m,n},y_{m,n})$, which according to
(\ref{sings}) is a singularity of the surface. Therefore $B_{m,n}$
is the conjugate $B$-cycle to the pinched $A$-cycle located at
$(x_{m,n},y_{m,n})$. These contours are shown in figure 3.

We can also rephrase the preceding paragraph in a way that leads
to a new insight about the ZZ branes. The association of the ZZ
branes with the singularities of the surface means that there is a
sense in which they are ``located" at the singularities. We can
make this more precise by recalling that the equations for the
singularities are the same as the ground ring relations. This
suggests that the ZZ branes and the ground ring are related in
some natural way. Indeed, one can show that the ZZ branes are {\it
eigenstates} of the ground ring elements, with eigenvalues
$(x_{m,n},y_{m,n})$:
\begin{equation}
\label{greigen}\begin{split}
 X|m,n\rangle &= x_{m,n}
 |m,n\rangle \cr
 Y|m,n\rangle &= y_{m,n}
 |m,n\rangle
\end{split}\end{equation}
We note in passing that this leads to a simple derivation of the
ring relations (\ref{grmult}) and (\ref{grrel}). More to the
point, however, (\ref{greigen}) makes precise the idea that the ZZ
branes are located at the singularities. According to
(\ref{greigen}), we can think of the ring generators $X$ and $Y$
as measuring the ``position" of the ZZ brane on $\CM_{p,q}$.

So far we have been considering a special closed-string background
corresponding to the Liouville action (\ref{LiouvilleS}) with the
cosmological constant interaction. This gives rise to the surface
described by (\ref{Fpq}). We can also consider more general
backgrounds, obtained by adding other physical operators to the
worldsheet action (e.g.\ the tachyons (\ref{tach})). These will
deform the equation of the Riemann surface, but in such a way as
to preserve its geometrical properties. In particular, the surface
will still be a finite-sheeted cover of the complex $x$ plane, and
it will still have a number of singularities. Thus the deformed
surface will still possess a uniformizing parameter $z$. In fact,
one can characterize the deformations as deformations of the
uniformizing map (\ref{Fpqsolz}):
\begin{equation}
 \label{deformations}
 \delta x = \epsilon R(z),\qquad \delta y = \epsilon S(z)
\end{equation}
with $R$ and $S$ polynomials in $z$. One can show that
infinitesimal deformations by closed string states correspond to
singularity preserving deformations of $\CM_{p,q}$ of the form
(\ref{deformations}). Conversely, the list of all polynomial
deformations to $x(z)$ and $y(z)$ captures the spectrum of
physical closed string states at all ghost numbers.

Of course, we can also imagine deformations of the surface which
do not preserve the singularities. These correspond to adding
$\CO\left({1 /g_{s}}\right)$ background ZZ branes. This has a nice
geometrical realization in terms of the contours of the surface:
the period
$\oint_{B_{m,n}} \omega$
creates $(m,n)$ ZZ branes, while the conjugate period
$\oint_{A_{m,n}} \omega$ measures how many are present:
\begin{equation}
\label{ZZcont}
\oint_{A_{m,n}} \omega = g_{s} N_{m,n}
\end{equation}
In target space, these deformations can be thought of as adding
background tachyons with the ``wrong'' Liouville dressing $\alpha
\ge {Q \over 2}$. Such tachyons diverge in the strong coupling
region $\phi\to +\infty$, and so they are naturally identified
with the addition of background ZZ branes \cite{KutasovFG}.

\begin{figure}
\begin{center}
\epsfig{file=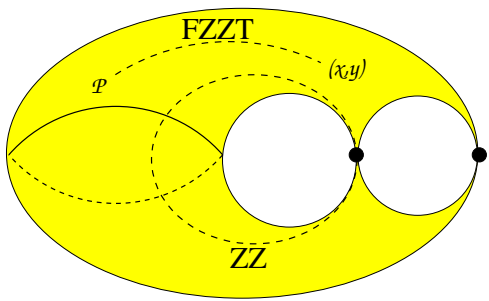, width=0.4\textwidth}
\end{center}
\caption{A Riemann surface with examples of FZZT and ZZ contours.}
\label{fig:contours}
\end{figure}

\section{Deriving the Dual Matrix Model}

Besides providing a unified description of minimal string theory,
the geometric picture outlined in the previous section has an
important added benefit: it leads directly to the dual matrix
model. The fact that minimal strings are dual to certain large $N$
random matrix models is well known (for a review and references,
see \cite{GinspargIS,DiFrancescoNW}), and the duality has been
verified in many different ways. Now, with our improved knowledge
of the worldsheet description of minimal string theory, we can
shed new light on this duality and basically derive it.

For simplicity, let us focus on the models with $(p=2,q=2k-1)$,
which are dual to the one matrix model
\begin{equation}
\label{onemat} Z(g) = \int dM\, e^{-{1\over g}\Tr\, V(M)}
\end{equation}
with $M$ an $N\times N$ Hermitian matrix. The surface for $p=2$ is
\begin{equation}
\label{surfaceex}
2 y^2 -1=T_{2k-1}(x)
\end{equation}
This describes a double cover of the complex $x$ plane on which
$y(x)$ is single valued. The two sheets are connected along a cut
$-\infty < x\le -1$. There are also $k$ singularities (pinched
cycles) located at
\begin{equation}
\label{singex}
 \bigg(x_{n}= \cos {2\pi n \over 2k-1}\ ,\
  y_{n}=0\bigg) \ , \quad n=1,...,k
\end{equation}

Now we can proceed to match the surface with quantities in the
matrix model. The discontinuity of $y(x)$ along the cut is the
eigenvalue density:
\begin{equation}
\label{eigendens}
 \rho(x)={\rm Im} \sqrt{2+2\,T_{2k-1}(x)}
\end{equation}
More generally, $y(x)$ corresponds to the force on an eigenvalue
(note that $y=0$ at the singularities), and the disk amplitude of
the FZZT brane
\begin{equation}
\label{Veff}
 Z(x)= \int^x y \, dx'=- {1\over
2} V_{eff}(x)
\end{equation}
is the effective potential of a probe eigenvalue. Since the ZZ
branes are located at the singularities, we conclude that they
correspond to eigenvalues at the stationary points of $V_{eff}(x)$
(where $y=0$). The cut at $x<-1$ corresponds to the Fermi sea --
the ZZ branes decay (condense) and fill the Fermi sea. The matrix
$M$ of the matrix model then corresponds to open strings between
$N \to \infty$ condensed ZZ branes.

The FZZT brane in the matrix model is described by the {\it
macroscopic loop operator}
\begin{equation}
\label{macroloop} W(x) = {\rm Tr}\, {\rm log}(x-M)
\end{equation}
Thus $Z(x)=\langle W(x)\rangle$, and $y=\partial_x Z(x)$ is the
resolvent of the matrix model. The full, nonperturbative FZZT
brane corresponds to worldsheets with any number of boundaries
(and handles). This is accomplished in the matrix model by
exponentiating $W(x)$, leading to a simple formula for the
FZZT-brane creation operator
\begin{equation}
\label{FZZTcreation}
\Psi(x) \sim \det(x-M)
\end{equation}
Note that we can write this as a Grassmann integral over $2N$
fermions $\chi_i$ and $\chi^\dagger_i$:
\begin{equation}
\label{FZZTferm} \det(x-M)= \int d\chi^\dagger d\chi\,
e^{\chi^\dagger(x-M) \chi}
\end{equation}
We interpret $\chi$, $\chi^\dagger$ to be {\it fermionic} open
strings between ZZ and FZZT branes.

\begin{figure}
\begin{center}
\epsfig{file=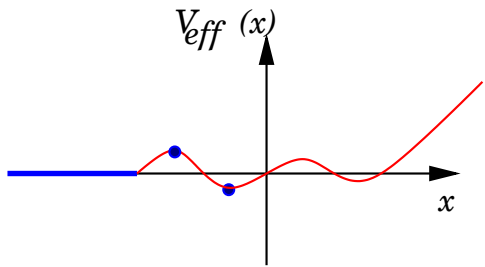, width=0.4\textwidth}
\end{center}
\caption{An example of the effective potential of the matrix
model. The blue line denotes the branch cut in the effective
potential (corresponding to the Fermi sea), and the blue dots at
the local extrema denote stationary eigenvalues corresponding to
ZZ branes.} \label{fig:veff}
\end{figure}

\section{Conclusions}

We have seen how an effective ``target space,'' consisting of a
certain Riemann surface $\CM_{p,q}$, emerges as the moduli space
of branes. This surface captures many of the properties of the
minimal string, including its D-branes, its spectrum of
closed-string operators, and their correlation functions. The
D-branes correspond to integrals of a certain one-form $\omega$ on
the Riemann surface, while the deformations of the surface encode
the closed-string observables (singularity preserving) and the
spectrum of localized branes (singularity destroying).

We also saw how this geometric picture is complemented by the
algebraic structure of the ground ring. In particular, the ring
relations controlled the correlation functions, the defining
equation of the surface, and the location of its singularities.

Finally, we gave a worldsheet derivation of the matrix model, and
added a new perspective to the understanding that the eigenvalues
of the matrix model are associated with D-branes
\cite{ShenkerUF,PolchinskiFQ,McGreevyKB,KlebanovKM}.

Let us conclude with the following comment on the regime of
validity of our results. Clearly, the geometrical picture
described here (which emerged from the disk amplitude of the FZZT
brane) is only meant to apply at the level of perturbation theory
in the string coupling. In fact, nonperturbative effects change
the ``target space" $\CM_{p,q}$ very dramatically
\cite{MaldacenaSN}. Order-by-order in perturbation theory, the
moduli space of FZZT branes is a multiple cover of the complex $x$
plane. But the exact FZZT observables are entire functions of $x$,
and therefore the exact moduli space is reduced to just a single
copy of the $x$ plane \cite{MaldacenaSN}.

\section*{Acknowledgements}
We would like to thank the organizers of Strings 2004 for their
hospitality and for the invitation to speak about our work. The
research of NS is supported in part by DOE grant
DE-FG02-90ER40542. The research of DS is supported in part by an
NSF Graduate Research Fellowship and by NSF grant PHY-0243680. Any
opinions, findings, and conclusions or recommendations expressed
in this material are those of the author(s) and do not necessarily
reflect the views of the National Science Foundation.

\end{document}